\documentclass[a4paper,fleqn,usenatbib]{mnras}

\usepackage{newtxtext,newtxmath}

\usepackage[T1]{fontenc}
\usepackage{ae,aecompl}

\usepackage{graphicx}	
\usepackage{amsmath}	
\usepackage{amssymb}	
\usepackage{amsbsy}
\usepackage{algorithm}
\usepackage{algorithmic}
\usepackage{bbm}
\usepackage{array}
\usepackage{mathrsfs}
\usepackage{float}
\usepackage{mathrsfs}
\usepackage[utf8]{inputenc}
\usepackage{mathtools}
\usepackage[dvipsnames,table]{xcolor}
\usepackage{lipsum}
\usepackage{pifont}
\usepackage{chngcntr}
\usepackage{hyperref}
\hypersetup{
    colorlinks = true,
    urlcolor = black,
    linkbordercolor = {white}
}
\usepackage[most]{tcolorbox}
\usepackage{etoolbox}
\usepackage{longtable}
\usepackage{pdflscape}
\usepackage{geometry}
\usepackage{enumitem}
\usepackage{booktabs}
\usepackage{multirow}

\makeatletter

\patchcmd{\NAT@citex}
  {\@citea\NAT@hyper@{%
     \NAT@nmfmt{\NAT@nm}%
     \hyper@natlinkbreak{\NAT@aysep\NAT@spacechar}{\@citeb\@extra@b@citeb}%
     \NAT@date}}
  {\@citea\NAT@nmfmt{\NAT@nm}%
   \NAT@aysep\NAT@spacechar\NAT@hyper@{\NAT@date}}{}{}

\patchcmd{\NAT@citex}
  {\@citea\NAT@hyper@{%
     \NAT@nmfmt{\NAT@nm}%
     \hyper@natlinkbreak{\NAT@spacechar\NAT@@open\if*#1*\else#1\NAT@spacechar\fi}%
       {\@citeb\@extra@b@citeb}%
     \NAT@date}}
  {\@citea\NAT@nmfmt{\NAT@nm}%
   \NAT@spacechar\NAT@@open\if*#1*\else#1\NAT@spacechar\fi\NAT@hyper@{\NAT@date}}
  {}{}

\makeatother

\AtBeginEnvironment{quote}{\singlespace\vspace{-\topsep}\small}
\AtEndEnvironment{quote}{\vspace{-\topsep}\endsinglespace}
\newif\ifblackandwhite
 \blackandwhitetrue
 \ifblackandwhite
  
\else

\fi

\newcommand {\sumin} {\sum_{i=1}^{n}}
\newcommand {\what} {\widehat}

\newcommand{\MYhref}[3][blue]{\href{#2}{\color{#1}{#3}}}%
\definecolor{block-gray}{gray}{0.98}
\newtcolorbox{blockquote}{colback=block-gray,grow to right by=-1mm,grow to left by=-1mm,boxrule=0pt,boxsep=0pt,breakable}

\title{Trend Filtering -- I. A Modern Statistical Tool for Time-Domain Astronomy and Astronomical Spectroscopy}
\author[C. A. Politsch et al.]{Collin A. Politsch,$^{1,2,3}$\thanks{E-mail: capolitsch@cmu.edu}
Jessi Cisewski-Kehe,$^{4}$
Rupert A. C. Croft,$^{3,5,6}$ \newauthor
and Larry Wasserman$^{1,2,3}$
\vspace{.2cm}
\\ 
$^{1}$ Department of Statistics \& Data Science, Carnegie Mellon University, Pittsburgh, PA 15213 \\
$^{2}$ Machine Learning Department, Carnegie Mellon University, Pittsburgh, PA 15213 \\
$^{3}$ McWilliams Center for Cosmology, Carnegie Mellon University, Pittsburgh, PA 15213 \\
$^{4}$ Department of Statistics and Data Science, Yale University, New Haven, CT 06520 \\
$^{5}$ Department of Physics, Carnegie Mellon University, Pittsburgh, PA 15213 \\
$^{6}$ School of Physics, University of Melbourne, VIC 3010, Australia
}

\date{Accepted XXX. Received YYY; in original form 2019 August 20}

\pubyear{2020}

\makeatletter
\def\BState{\State\hskip-\ALG@thistlm}
\makeatother

\begin{document}
\label{firstpage}
\pagerange{\pageref{firstpage}--\pageref{lastpage}}
\maketitle
\title{Trend Filtering in Astronomy}
\begin{abstract}

The problem of denoising a one-dimensional signal possessing varying degrees of smoothness is ubiquitous in time-domain astronomy and astronomical spectroscopy. For example, in the time domain, an astronomical object may exhibit a smoothly varying intensity that is occasionally interrupted by abrupt dips or spikes. Likewise, in the spectroscopic setting, a noiseless spectrum typically contains intervals of relative smoothness mixed with localized higher frequency components such as emission peaks and absorption lines. In this work, we present trend filtering, a modern nonparametric statistical tool that yields significant improvements in this broad problem space of denoising \emph{spatially heterogeneous} signals. When the underlying signal is spatially heterogeneous, trend filtering is superior to any statistical estimator that is a linear combination of the observed data---including kernel smoothers, LOESS, smoothing splines, Gaussian process regression, and many other popular methods. Furthermore, the trend filtering estimate can be computed with practical and scalable efficiency via a specialized convex optimization algorithm, e.g. handling sample sizes of $n\gtrsim10^7$ within a few minutes. In a companion paper, we explicitly demonstrate the broad utility of trend filtering to observational astronomy by carrying out a diverse set of spectroscopic and time-domain analyses.


\end{abstract}




\section{Introduction}
\label{sec:1}
Many astronomical observations produce one-dimensional data with varying (or unknown) degrees of smoothness. These include data from time-domain astronomy, where transient events such as supernovae can show light-curve variations on timescales ranging from seconds to years \citep[e.g.,][]{Dim2017,Tolstov2019}. Similarly, in astronomical spectroscopy, with wavelength (or frequency) as the input variable, sharp absorption or emission-line features can be present alongside smoothly varying black-body or other continuum radiation \citep[see, e.g.,][]{Tennyson2019}. In each of these general settings, we observe a signal plus noise and would like to denoise the signal as accurately as possible. Indeed the set of statistical tools available for addressing this general problem is quite vast. \linebreak Commonly used nonparametric regression methods include kernel smoothers \citep[e.g.,][]{hall2002,Rupert1}, local polynomial regression \citep[LOESS; e.g.,][]{Maron_2003,Persson_2004}, splines \citep[e.g.,][]{Peiris_2009,Contreras_2010,dhawan15}, Gaussian process \linebreak regression (e.g., \citealt{Gibson_2012,Aigrain_2016,G_mez_Valent_2018}), and wavelet decompositions \citep[e.g.,][]{Fligge1997,Theuns_2000,golkhou14}. A rich and elegant statistical literature exists on the theoretical and practical achievements of these methods (see, e.g., \citealt{Gyorfi}; \citealt{Wasserman}; \citealt{ESL} for general references). However, when the underlying signal is \emph{spatially heterogeneous}, i.e. exhibits varying degrees of smoothness, the power of classical statistical literature is quite limited. Kernels, LOESS, smoothing splines, and Gaussian process regression belong to a broad family of nonparametric methods called \emph{linear smoothers}, which has been shown to be uniformly suboptimal for estimating spatially heterogeneous signals \cite[][]{Nemirovskii_1985, Nemirovskii_1985b, Donoho}. The common limitation of these methods is that they are not locally adaptive; i.e., by construction, they do not adapt to local degrees of smoothness in a signal. In particular, continuing with the example of a smoothly varying signal with occasional sharp features, a linear smoother will tend to oversmooth the sharp features and/or overfit the smooth regions in its effort to optimally balance statistical bias and variance. Considerable effort has been made to address this problem by locally varying the hyperparameter(s) of a linear smoother. For example, locally varying the kernel bandwidth \cite[e.g.,][]{muller1987,Fan1992b,Fan_1995,Lepskii1997OptimalSA,Gijbels_1998} irregularly varying spline knot locations \cite[e.g.,][]{de1974good,jupp1978approximation,10.1093/biomet/88.4.1055}, and constructing non-stationary covariance functions for Gaussian process regression \cite[e.g.,][]{10.2307/3647549,paciorek2004nonstationary,Schervish2}. However, since hyperparameters typically need to be estimated from the data, such exponential increases in the hyperparameter complexity severely limit the practicality of choosing the hyperparameters in a fully data-driven, generalizable, and computationally efficient fashion. Wavelet decompositions offer an elegant solution to the problem of estimating spatially heterogeneous signals, providing both statistical optimality \cite[e.g.,][]{Donoho_1994b,Donoho} and only requiring data-driven tuning of a single (scalar) hyperparameter. Wavelets, however, possess the practical limitation of requiring a stringent analysis setting, e.g. equally-spaced inputs and sample size equal to a power of two, among other provisions; and when these conditions are violated, the optimality guarantees are void. So, seemingly at an impasse, the motivating question for this work is \emph{can we have the best of both worlds?} More precisely, is there a statistical tool that simultaneously possesses the following properties:
\begin{enumerate}[leftmargin=*,labelindent=12pt,label={\bf P\arabic*.},ref=P\arabic*]
\item Statistical optimality for estimating spatially heterogeneous signals \label{P1}
\item Practical analysis assumptions; for example, not limited to equally-spaced inputs \label{P2}
\item Practical and scalable computational speed \label{P3}
\item A one-dimensional hyperparameter space, with automatic data-driven methods for selection \label{P4}
\end{enumerate}
In this paper we introduce trend filtering \citep{Tibs}, a statistical method that is new to the astronomical literature and provides a strong affirmative answer to this question. 

The layout of this paper is as follows. In Section \ref{sec:background} we provide both theoretical and empirical evidence of the superiority of trend filtering for estimating spatially heterogeneous signals compared to classical statistical methods. In Section \ref{sec:TF} we introduce trend filtering, including a general overview of the estimator's machinery, its connection to spline methods, automatic methods for choosing the hyperparameter, uncertainty quantification, generalizations, and recommended software implementations in various programming languages. In \cite{Politsch_2020b}---hereafter referred to as Paper~II---we directly illustrate the broad utility of trend filtering to astronomy by conducting various analyses of spectra and light curves.

\section{Classical statistical methods and their limitations}
\label{sec:background}

We begin this section by providing background and motivation for the nonparametric approach to estimating (or denoising) signals. We then discuss statistical optimality for estimating spatially heterogeneous signals, with an emphasis on providing evidence for the claim that trend filtering is superior to classical statistical methods in this highly general setting. Finally, we end this section by illustrating this superiority with a direct empirical comparison of trend filtering and several popular classical methods on simulated observations of a spatially heterogeneous signal.

\subsection{Nonparametric regression}
\label{subsec:nonpar}

Suppose we observe noisy measurements of a response variable of interest (e.g., flux, magnitude, photon counts) according to the data generating process (DGP)
\begin{equation}
f(t_i) = f_0(t_i) + \epsilon_i,  \hfill i=1,\dots,n \label{observational model}
\end{equation} 
where $f_0(t_i)$ is the signal at input $t_i$ (e.g., a time or wavelength) and $\epsilon_i$ is the noise at $t_i$ that contaminates the signal, giving rise to the observation $f(t_i)$. Let $t_1,\dots,t_n\in(a,b)$ denote the observed input interval and $\mathbb{E}[\epsilon_i] = 0$ (where we use $\mathbb{E}[\cdot]$ to denote mathematical expectation). Here, the general statistical problem is to estimate (or \emph{denoise}) the underlying signal $f_0$ from the observations as accurately as possible. In the nonparametric setting, we refrain from making strong \emph{a priori} assumptions about $f_0$ that could lead to significant modeling bias, e.g. assuming a power law or a light-curve/spectral template fit. Mathematically, a nonparametric approach is defined through the deliberately weak assumption $f_0 \in \mathcal{F}$ (i.e. the signal belongs to the function class $\mathcal{F}$) where $\mathcal{F}$ is \emph{infinite-dimensional}. In other words, the assumed class of all possible signals $\mathcal{F}$ cannot be spanned by a finite number of parameters. Contrast this to the assumption that the signal follows a $p$th degree power law, i.e. $f_0\in\mathcal{F}_{\text{PL}}$ where
\begin{equation}
\mathcal{F}_{\text{PL}} = \Bigg\{f_0 : f_0(t) = \beta_0 + \sum_{j=1}^{p}\beta_j t^{j}\Bigg\}, \label{FPL}
\end{equation} 
a class that is spanned by $p+1$ parameters. Similarly, given a set of $p$ spectral/light-curve templates $b_1(t),\dots,b_p(t)$, the usual template-fitting assumption is that $f_0\in\mathcal{F}_{\text{TEMP}}$ where\begin{equation}
\mathcal{F}_{\text{TEMP}} = \Bigg\{f_0 : f_0(t) = \beta_0 + \sum_{j=1}^{p}\beta_j b_j((t-s)/v)\Bigg\} \label{FTEMP}
\end{equation}
and $s$ and $v$ are horizontal shift and scale hyperparameters, respectively. 
Both (\ref{FPL}) and (\ref{FTEMP}) represent very stringent assumptions about the underlying signal $f_0$. If the signal is anything other than exactly a power law in $t$---a highly unlikely occurrence---nontrivial statistical bias will arise by modeling it as such. Likewise, if a class of signals has a rich physical diversity (e.g., Type Ia supernova light curves; \citealt{Woosley_2007}) that is not sufficiently spanned by the library of templates used in modeling, then statistical biases will arise. Depending on the size of the imbalance between class diversity and the completeness of the template basis, the biases could be significant. Moreover, these biases are rarely tracked by uncertainty quantification. To be clear, this is not a uniform criticism of template-fitting. For example, templates are exceptionally powerful tools for object classification and redshift estimation (e.g., \citealt{Howell_2005}, \citealt{Bolton_2012}). Furthermore, much of our discussion in Paper~II centers around utilizing the flexible nonparametric nature of trend filtering to construct more complete spectral/light-curve template libraries for various observational objects and transient events.


Let $\what{f}_0$ be any statistical estimator for the signal $f_0$, derived from the noisy observations in (\ref{observational model}). Further, let $p_t(t)$ denote the probability density function (pdf) that specifies the sampling distribution of the inputs on the interval $(a,b)$, and let $\sigma^{2}(t) = \text{Var}(\epsilon(t))$ denote the noise level at input $t$. In order to assess the accuracy of the estimator it is common to consider the mean-squared prediction error (MSPE):
\begin{align}
R(\what{f}_0) &= \mathbb{E}\big[(\what{f}_0 - f)^2\big] \\
&= \mathbb{E}\big[(\what{f}_0 - f_0)^2\big] + \overline{\sigma}^{2} \label{bv2} \\
&= \int_{a}^{b}\Big(\text{Bias}^2(\what{f}_0(t)) + \text{Var}(\what{f}_0(t))\Big)\cdot p_{t}(t)dt + \overline{\sigma}^{2},\label{biasvariancedecomp}
\end{align}
where
\begin{align}
\text{Bias}(\what{f}_0(t)) &= \mathbb{E}[\what{f}_0(t)] - f_0(t) \\
\text{Var}(\what{f}_0(t)) &= \mathbb{E}\Big(\what{f}_0(t) - \mathbb{E}[\what{f}_0(t)]\Big)^2 \\
\overline{\sigma}^2 &= \int_{a}^{b} \sigma^{2}(t)\cdot p_{t}(t)dt. \label{mean_noise}
\end{align}
The equality in (\ref{biasvariancedecomp}) is commonly referred to as the bias-variance decomposition. The first term is the squared bias of the estimator $\what{f}_0$ (integrated over the input interval) and intuitively measures how appropriate the chosen statistical estimator is for modeling the observed phenomenon. The second term is the variance of the estimator that measures how stable or sensitive the estimator is to the observed data. And the third term is the irreducible error---the minimum prediction error we cannot hope to improve upon. The bias-variance decomposition therefore illustrates that an optimal estimator is one that combines appropriate modeling assumptions (low bias) with high stability (low variance).

\subsubsection{Statistical optimality (minimax theory)}
\label{subsubsec:minimax}

In this section, we briefly discuss a mathematical framework for evaluating the performance of statistical methods over nonparametric signal classes in order to demonstrate that the superiority of trend filtering is a highly general result. Ignoring the irreducible error, the problem of minimizing the MSPE of a statistical estimator can be equivalently stated as a minimization of the first term in (\ref{bv2})---the mean-squared estimation error (MSEE). In practice, low bias is attained by only making very weak assumptions about what the underlying signal may look like, e.g. $f_0$ has $k$ continuous derivatives. An ideal statistical estimator for estimating signals in such a class (call it $\mathcal{F}$) may then be defined as
\begin{equation}
\inf_{\what{f}_0} \Big(\sup_{f_0\in\mathcal{F}}\mathbb{E}\big[(\what{f}_0 - f_0)^2\big]\Big). \label{minimax}
\end{equation}
That is, we would like our statistical estimator to be the minimizer (infimum) of the worst-case (supremum) MSEE over the signal class $\mathcal{F}$. This is rarely a mathematically tractable problem for any practical signal class $\mathcal{F}$. A more tractable approach is to consider how the worst-case MSEE behaves as a function of the sample size $n$. A reasonable baseline metric for a statistical estimator is to require that it satisfies
\begin{equation}
\sup_{f_0\in\mathcal{F}}\mathbb{E}\big[(\what{f}_0 - f_0)^2\big]\rightarrow0
\end{equation}
as $n\rightarrow\infty$. That is, for any signal $f_0\in\mathcal{F}$, when a large amount of data is available, $\what{f}_0$ gets arbitrarily close to the true signal. In any practical situation, this is not true for parametric models because the bias component of the decomposition never vanishes. This, however, is a widely-held---perhaps, defining---property of nonparametric methods. Therefore, in order to distinguish optimality among nonparametric estimators, we require a stronger metric. In particular, we study \emph{how quickly} the worst-case error goes to zero as more data are observed. This is the core idea of a rich area of statistical literature called \emph{minimax theory} \cite[see, e.g.,][]{vaart_1998,Wasserman,Tsybakov_2008}. For many infinite-dimensional classes of signals, theoretical lower-bounds exist on the rate at which the MSEE of \emph{any} statistical estimator can approach zero. Therefore, if a statistical estimator is shown to achieve that rate, it can be considered optimal for estimating that class of signals. Formally, letting $g(n)$ be the rate at which the MSEE of the theoretically optimal estimator (\ref{minimax}) goes to zero (a monotonically decreasing function in $n$),  we would like our estimator $\what{f}_0$ to satisfy
\begin{equation}
\sup_{f_0\in\mathcal{F}}\mathbb{E}\big[(\what{f}_0 - f_0)^2\big] = \mathcal{O}(g(n)),
\end{equation}
where we use $\mathcal{O}(\cdot)$ to denote big O notation. If this is shown to be true, we say the estimator \emph{achieves the minimax rate over the signal class $\mathcal{F}$}. Loosely speaking, we are stating that a \emph{minimax optimal} estimator is an estimator that learns the signal from the data just as quickly as the theoretical gold standard estimator (\ref{minimax}).


\subsubsection{Spatially heterogeneous signals}
\label{subsubsec:hetero}

Thus far we have only specified that the signal underlying most one-dimensional astronomical observations should be assumed to belong to a class $\mathcal{F}$ that is infinite-dimensional (i.e. nonparametric). Further, in Section \ref{subsubsec:minimax} we introduced the standard metric used to measure the performance of a statistical estimator over an infinite-dimensional class of signals. Recalling the discussion in the abstract and Section \ref{sec:1}, trend filtering provides significant advances for estimating signals that exhibit varying degrees of smoothness across the input domain. We restate this definition below.

\vspace{0.2cm}

\begin{blockquote}
{\bf Definition.} A \emph{spatially heterogeneous} signal is a signal that exhibits varying degrees of smoothness in different regions of its input domain. \\

{\bf Example.} A smooth light curve with abrupt transient events. \\

{\bf Example.} An electromagnetic spectrum with smooth continuum radiation and sharp absorption/emission-line features.
\end{blockquote}
\vspace{0.2cm}

\noindent To complement the above definition we may also loosely define a \emph{spatially homogeneous signal} as a signal that is \emph{either} smooth \emph{or} wiggly\footnote{This is, in fact, a technical term used in the statistical literature.} across its input domain, but not both. As ``smoothness'' can be quantified in various ways these definitions are intentionally mathematically imprecise. A class that is commonly considered in the statistical literature is the $L_2$ Sobolev class:
\begin{equation}
\mathcal{F}_{2,k}(C_1) := \Bigg\{f_0: \int_{a}^{b}f_0^{(k)}(t)^2dt < C_1 \Bigg\}, \hfill C_1 > 0,\;k\in\mathbb{N}. \label{l2sobo}
\end{equation}
That is, an $L_2$ Sobolev class is a class of all signals such that the integral of the square (the ``$L_2$ norm'') of the $k$th derivative of each signal is less than some constant $C_1$. Statistical optimality in the sense of Section \ref{subsubsec:minimax} for estimating signals in these classes (and some other closely related ones) is widely held among statistical methods in the classical toolkit; for example, kernel smoothers \cite[][]{Ibragimov_1980,stone1982}, LOESS \cite[][]{fan1993,Fan1997}, and smoothing splines \cite[][]{Nussbaum}. However, a seminal result by \cite{Nemirovskii_1985} and \cite{Nemirovskii_1985b} showed that a statistical estimator can be minimax optimal over signal classes of the form (\ref{l2sobo}) and still perform quite poorly on other signals. In particular, the authors showed that, when considering the broader $L_1$ Sobolev class\begin{equation}
\mathcal{F}_{1,k}(C_2) := \Bigg\{f_0: \int_{a}^{b}\big|f_0^{(k)}(t)\big|dt < C_2 \Bigg\}, \hfill C_2 > 0,\;k\in\mathbb{N},
\label{l1sobo}
\end{equation}
all linear smoothers\footnote{A \emph{linear smoother} is a statistical estimator that is a linear combination of the observed data. Many popular statistical estimators, although often motivated from seemingly disparate premises, can be shown to fall under this definition. See, e.g., \cite{Wasserman} for more details.}---including kernels, LOESS, smoothing splines, Gaussian process regression, and many other methods---are strictly suboptimal. The key difference between these two types of classes is that $L_2$ Sobolev classes are rich in spatially homogeneous signals but not spatially heterogeneous signals, while $L_1$ Sobolev classes\footnote{The $L_1$ Sobolev class is often generalized to a nearly equivalent but slightly larger class---namely, signals with derivatives of bounded variation. See \cite{Tibs} for the generalized definition.} are rich in both \cite[see, e.g.,][]{Donoho}. 

The intuition of this result is that linear smoothers cannot optimally recover signals that exhibit varying degrees of smoothness across their input domain because they operate as if the signal possesses a fixed degree of smoothness. For example, this intuition is perhaps most clear when considering a kernel smoother with a fixed bandwidth. The result of \cite{Nemirovskii_1985} and \cite{Nemirovskii_1985b} therefore implies that, in order to achieve statistical optimality for estimating spatially heterogeneous signals, a statistical estimator must be nonlinear (more specifically, it must be locally adaptive). \cite{Tibs} showed that trend filtering is minimax optimal for estimating signals in $L_1$ Sobolev classes. Since $L_2$ Sobolev classes are contained within $L_1$ Sobolev classes, this result also guarantees that trend filtering is also minimax optimal for estimating signals in $L_2$ Sobolev classes. Wavelets share this property, but require restrictive assumptions on the sampling of the data \cite[][]{Donoho_1994b}. 

\emph{How large is this performance gap?} The collective results of \cite{Nemirovskii_1985}, \cite{Nemirovskii_1985b}, and \cite{Tibs} reveal that the performance gap between trend filtering and linear smoothers when estimating spatially heterogeneous signals is significant. For example, when $k=0$, the minimax rate over $L_1$ Sobolev classes (which trend filtering achieves) is $n^{-2/3}$, but linear smoothers cannot achieve better than $n^{-1/2}$. 
To put this in perspective, this result says that the trend filtering estimator, training on $n$ data points, learns these signals with varying smoothness as quickly as a linear smoother training on $n^{4/3}$ data points. As we demonstrate in the next section, this gap in theoretical optimality has clear practical consequences. 

In order to minimize the pervasion of technical statistical jargon throughout the paper, henceforth we simply refer to a statistical estimator that achieves the minimax rate over $L_2$ Sobolev classes as \emph{statistically optimal for estimating spatially homogeneous signals}, and we refer to a statistical estimator that achieves the minimax rate over $L_1$ Sobolev classes as \emph{statistically optimal for estimating spatially heterogeneous signals}. As previously mentioned, the latter implies the former, but not vice versa.

\subsection{Empirical comparison}
\label{subsec:comparison}

In this section we analyze noisy observations of a simulated spatially heterogeneous signal in order to compare the empirical performance of trend filtering and several classical statistical methods---namely, LOESS, smoothing splines, and Gaussian process regression. The mock observations are simulated on an unequally-spaced grid $t_1,\dots,t_n \sim \text{Unif}(0,1)$ according to the data generating process
\begin{equation}
f(t_i) = f_0(t_i) + \epsilon_i \label{DGP2_toy}
\end{equation}
with
\begin{equation}
f_0(t_i) = 6 \sum_{k=1}^{3}(t_i - 0.5)^k + 2.5\sum_{j=1}^{4}(-1)^{j}\phi_j(t_i),
\end{equation}
where $\phi_j(t),$ $j=1,\dots,4$ are compactly-supported radial basis functions distributed throughout the input space and $\epsilon_i \sim N(0,0.125^2)$. We therefore construct the signal $f_0$ to have a smoothly varying global trend with four sharp localized features---two dips and two spikes. The signal and noisy observations are shown in the top panel of Figure~\ref{spline}. 

\begin{figure*}
\centering
\includegraphics[width = \textwidth]{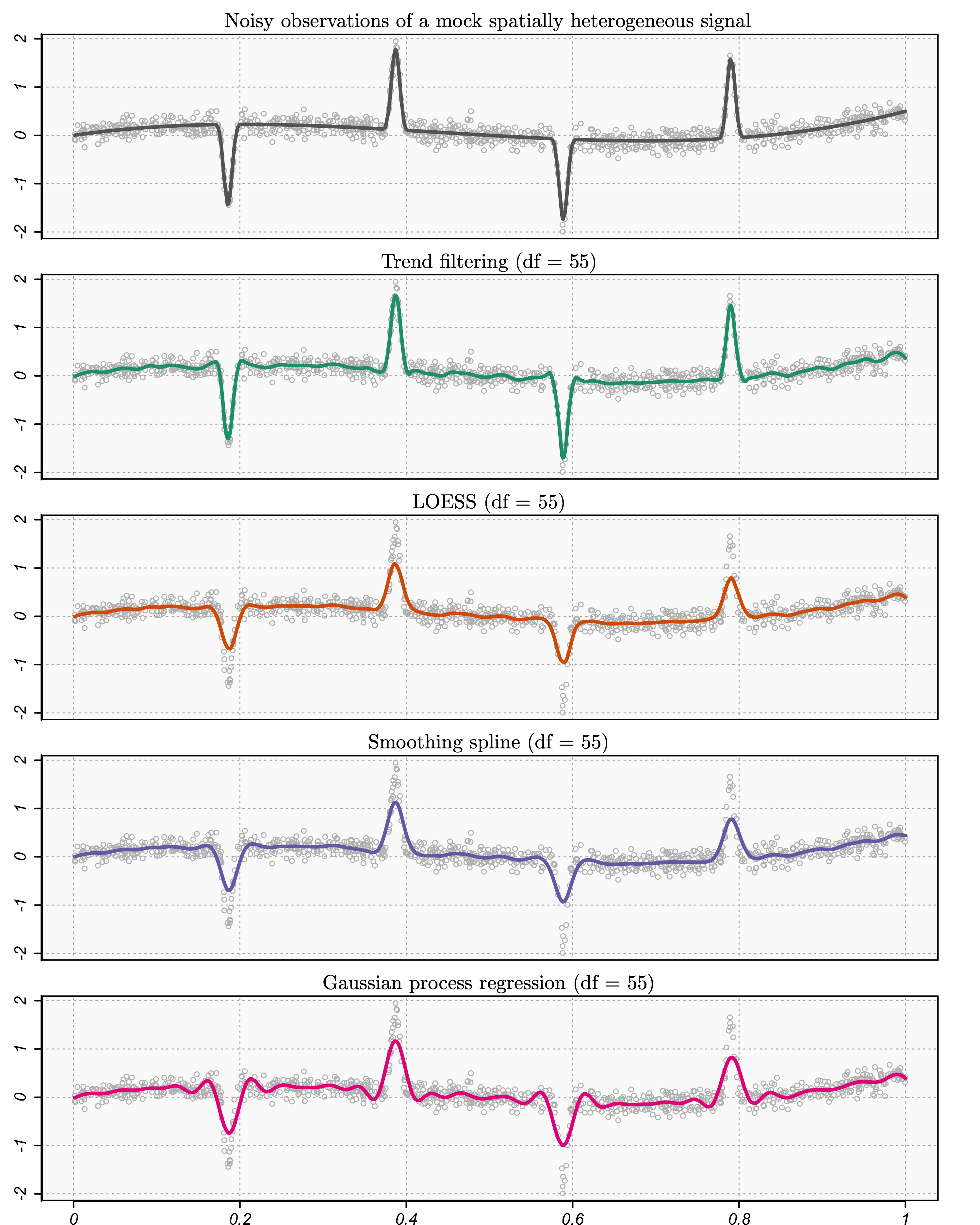} \\
\caption{Comparison of statistical methods on data simulated from a spatially heterogeneous signal. Each statistical estimator is fixed to have 55 effective degrees of freedom in order to facilitate a direct comparison. The trend filtering estimator is able to sufficiently distribute its effective degrees of freedom such that it simultaneously recovers the smoothness of the global trend, as well as the abrupt localized features. The LOESS, smoothing spline, and Gaussian process regression each estimates the smooth global trend reasonably well here, but significantly oversmooths the sharp peaks and dips. Here, we utilize quadratic trend filtering (see Section~\ref{subsec:TFdef}).}
\label{spline}
\end{figure*}
\begin{figure*}
\centering
\includegraphics[width = \textwidth]{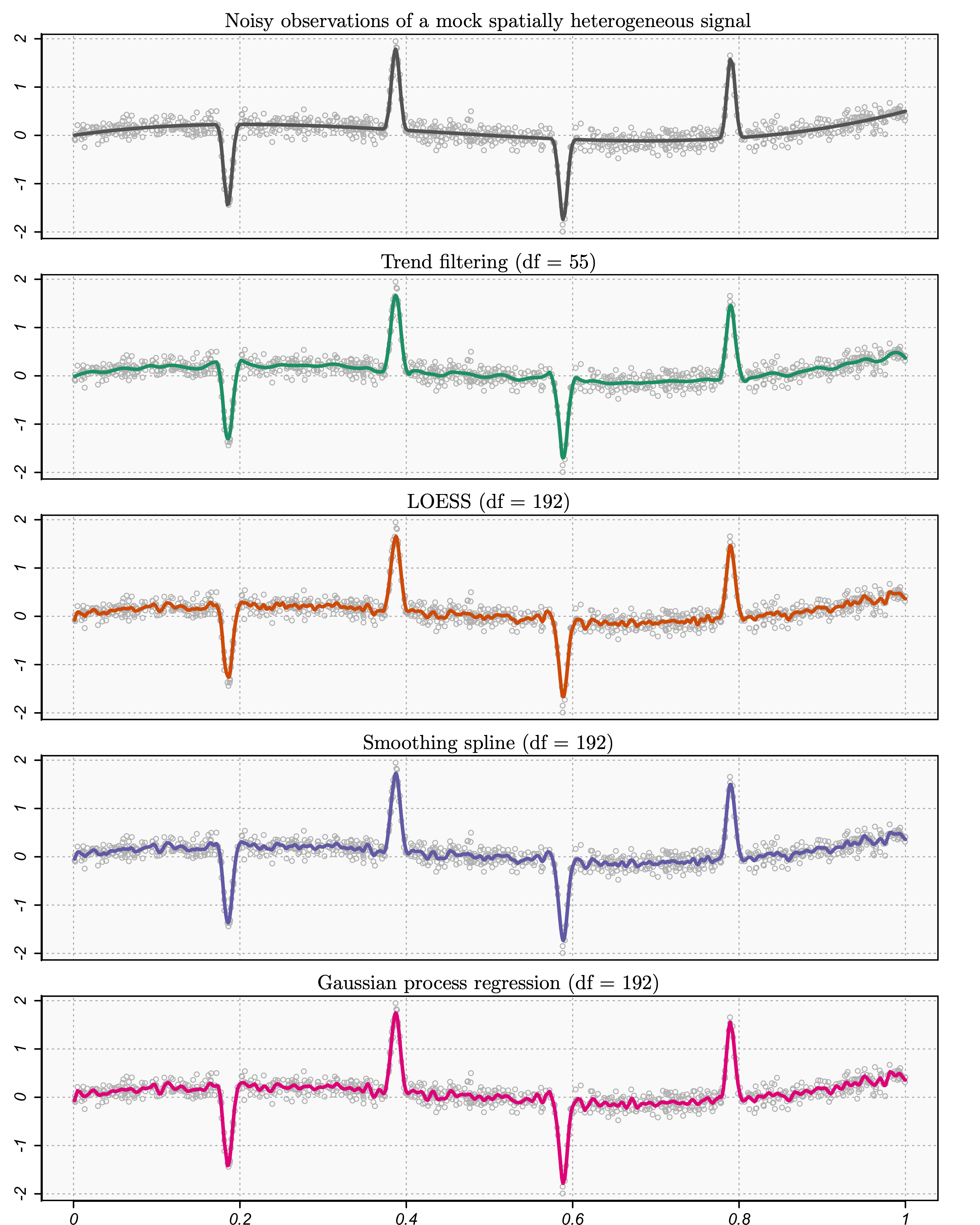} \\
\caption{(Continued): Comparison of statistical methods on data simulated from a spatially heterogeneous signal. Here, each of the linear smoothers (i.e. the LOESS, smoothing spline, and Gaussian process regression) is fixed at 192 effective degrees of freedom---the complexity necessary for each estimator to recover the sharp localized features approximately as well as the trend filtering estimator with 55 effective degrees of freedom. While the linear smoothers now estimate the four abrupt features well, each severely overfits the data in the other regions of the input domain.}
\label{spline2}
\end{figure*}

In order to facilitate the comparison of methods we utilize a metric for the total statistical complexity (i.e. total wiggliness) of an estimator known as the \emph{effective degrees of freedom}  \cite[see, e.g.,][]{tibshirani2015degrees}. Formally, the effective degrees of freedom of an estimator $\what{f}_0$ is defined as
\begin{equation}
\text{df}(\what{f}_0) = \overline{\sigma}^{\;-2}{\sumin \text{Cov}(\what{f}_0(t_i), f(t_i))} \label{edf}
\end{equation} 
where $\overline{\sigma}^{2}$ is defined in (\ref{mean_noise}). In Figure~\ref{spline} we fix all estimators to have 55 effective degrees of freedom. This exercise provides insight into how each estimator relatively distributes its complexity across the input domain. In the second panel we see that the trend filtering estimate has sufficiently recovered the underlying signal, including both the smoothness of the global trend and the abruptness of the localized features. All three of the linear smoothers, on the other hand, severely oversmooth the localized peaks and dips. Gaussian process regression also exhibits some undesirable oscillatory features that do not correspond to any real trend in the signal. In order to better recover the localized features the linear smoothers require a more complex fit, i.e. smaller LOESS kernel bandwidth, smaller smoothing spline penalization, and smaller Gaussian process noise-signal variance. In Figure~\ref{spline2} we show the same comparison, but we grant the linear smoothers more complexity. Specifically, in order to recover the sharp features comparably with the trend filtering estimator with 55 effective degrees of freedom, the linear smoothers require 192 effective degrees of freedom---approximately 3.5 times the complexity. As a result, although they now adequately recover the peaks and dips, each linear smoother severely overfits the data in the other regions of the input domain, resulting in many spurious fluctuations.

As discussed in Section~\ref{subsubsec:hetero}, the suboptimality of LOESS, smoothing splines, and Gaussian process regression illustrated in this example is an inherent limitation of the broad \emph{linear smoother} family of statistical estimators.
Linear smoothers are adequate tools for estimating signals that exhibit approximately the same degree of smoothness throughout their input domain. 
However, when a signal is expected to exhibit varying degrees of smoothness across its domain, a locally-adaptive statistical estimator is needed.

\section{Trend filtering}
\label{sec:TF}

Trend filtering, in its original form, was independently proposed in the computer vision literature (\citealt{steidl}) and the applied mathematics literature (\citealt{trendfilter}), and has recently been further developed in the statistical and machine learning literature, most notably with \cite{genlasso}, \cite{Tibs}, \cite{Yuxiang}, and \cite{Ramdas}. This work is in no way related to the work of \cite{TFA}, which goes by a similar name. At a high level, trend filtering is closely related to two familiar nonparametric regression methods: variable-knot regression splines and smoothing splines. We elaborate on these relationships below.

\subsection{Closely-related methods}
\label{subsec:related}

Splines have long played a central role in estimating complex signals (see, e.g., \citealt{Boor1978APG} and \citealt{wahba1990spline} for general references). Formally, a $k$th order spline is a piecewise polynomial (i.e. piecewise power law) of degree $k$ that is continuous and has $k-1$ continuous derivatives at the knots. As their names suggest, variable-knot regression splines and smoothing splines center around fitting splines to observational data. Recall from (\ref{observational model}) the observational data generating process (DGP)
\begin{equation}
f(t_i) = f_0(t_i) + \epsilon_i,  \hfill t_1,\dots,t_n\in(a,b), \label{observational_model2}
\end{equation} 
where $f(t_i)$ is a noisy measurement of the signal $f_0(t_i)$, and $\mathbb{E}[\epsilon_i] = 0$. Given a set of knots $\kappa_1,\dots,\kappa_p\in(a,b)$, the space of all $k$th order splines on the interval $(a,b)$ with knots at $\kappa_1,\dots,\kappa_p$ can be parametrized via a basis representation
\begin{equation}
m(t) = \sum_j \beta_j \eta_j(t),
\end{equation}
where $\{\eta_j\}$ is typically the truncated power basis or B-spline basis. A suitable estimator for the signal $f_0$ may then be
\begin{equation}
\what{f}_0(t) = \sum_j \what{\beta}_j \eta_j(t),
\end{equation}
where the $\what{\beta}_j$ are the ordinary least-squares (OLS) estimates of the basis coefficients. This is called a \emph{regression spline}. The question of course remains where to place the knots. 

\subsubsection{Variable-knot regression splines}
\label{subsec:variableknot}

The variable-knot (or free-knot) regression spline approach is to consider all regression spline estimators with knots at a subset of the observed inputs, i.e. $\{\kappa_1,\dots,\kappa_p\}\subset \{t_1,\dots,t_n\}$ for all possible $p$. Formally, the variable-knot regression spline estimator is the solution to the following constrained least-squares minimization problem:
\begin{equation}
\begin{aligned}
\min_{\{\beta_j\}} \quad  &\sum_{i=1}^{n}\Bigg(f(t_i)- \sum_{j}\beta_j\eta_j(t_i)\Bigg)^2\\
\textrm{s.t.} & \quad  \sum_{j\geq k+2}\mathbbm{1}\{\beta_j\neq 0\} = p \\
& \quad \;p \geq 0
\end{aligned} \label{variableknot}
\end{equation}
where $p\geq0$ is the number of knots in the spline and $\mathbbm{1}(\cdot)$ is the indicator function satisfying
\begin{equation}
\mathbbm{1}\{\beta_j\neq 0\} = \begin{cases}
1 & \beta_j\neq 0, \\
0 & \beta_j= 0.
\end{cases}
\end{equation}
Furthermore, note that the equality constraint on the basis coefficients excludes those of the ``first'' $k+1$ basis functions that span the space of global polynomials and only counts the number of active basis functions that produce knots. The variable-knot regression spline optimization is therefore a problem of finding the \emph{best subset} of knots for the regression spline estimator. Due to the sparsity of the coefficient constraint, the variable-knot regression spline estimator allows for highly locally-adaptive behavior for estimating signals that exhibit varying degrees of smoothness. However, the problem itself cannot be solved in polynomial time, requiring an exhaustive combinatorial search over all $\sim$2$^n$ feasible models. It is common to utilize stepwise procedures based on iterative addition and deletion of knots in the active set, but these partial searches over the feasible set inherently provide no guarantee of finding the optimal global solution to (\ref{variableknot}).

In order to make the connection to trend filtering more explicit it is helpful to reformulate the constrained minimization~(\ref{variableknot}) into the following penalized unconstrained minimization problem:
\begin{equation}
\min_{\{\beta_j\}} \quad  \sum_{i=1}^{n}\Bigg(f(t_i)- \sum_{j}\beta_j\eta_j(t_i)\Bigg)^2 + \gamma \sum_{j\geq k+2}\mathbbm{1}\{\beta_j\neq 0\}, \label{L0}
\end{equation}
where $\gamma>0$ is a hyperparameter that determines the number of knots in the spline and the sum of indicator functions serves as a smoothness ``penalty'' on the ordinary least-squares minimization. Penalized regression is a popular area of statistical methodology \cite[see, e.g.,][]{ESL}, in which the cost functional (i.e. the quantity to be minimized) quantifies a tradeoff between the training error of the estimator (here, the sum of squared residuals) and the statistical complexity of the estimator (here, the number of knots in the spline). In particular, (\ref{L0}) is known as an $\ell_0$-penalized least-squares regression because of the penalty's connection to the mathematical $\ell_0$ vector quasi-norm.

\subsubsection{Smoothing splines}

Smoothing splines counteract the computational issue faced by variable-knot regression splines by simply placing knots at all of the observed inputs $t_1,\dots,t_n$ and regularizing the smoothness of the fitted spline. For example, letting $\mathcal{G}$ be the space of all cubic natural splines with knots at $t_1,\dots,t_n$, the cubic smoothing spline estimator is the solution to the optimization problem
\begin{equation}
\min_{m \in \mathcal{G}} \quad  \sum_{i=1}^{n}\big(f(t_i)- m(t_i)\big)^2 + \gamma \int_{a}^{b}\big(m''(t)\big)^2dt, \label{smoothspline1}
\end{equation}
where $m''$ is the second derivative of $m$ and $\gamma>0$ tunes the amount of regularization. Letting $\eta_1,\dots,\eta_n$ be a basis for cubic natural splines with knots at the observed inputs, (\ref{smoothspline1}) can be equivalently stated as a minimization over the basis coefficients:
\begin{equation}
\min_{\{\beta_j\}} \quad  \sum_{i=1}^{n}\Bigg(f(t_i)- \sum_{j}\beta_j\eta_j(t_i)\Bigg)^2 + \gamma \sum_{j,k=1}^n\beta_j\beta_k\omega_{jk} \label{L2}
\end{equation}
where
\begin{equation}
\omega_{jk} = \int_{a}^{b}\eta_j''(t)\eta_k''(t)dt.
\end{equation}
The cost functional~(\ref{L2}) is differentiable and leads to a linear system with a special sparse structure (i.e. bandedness), which yields a solution that can both be found in closed-form and computed very quickly---in $O(n)$ elementary operations. This particular choice of cost functional, however, produces an estimator that is a linear combination of the observations---a \emph{linear smoother}. Therefore, as discussed and demonstrated in Section \ref{sec:background}, smoothing splines are suboptimal for estimating spatially heterogeneous signals. Equation~(\ref{L2}) is known as an $\ell_2$-penalized least-squares regression because of the penalty's connection to the mathematical $\ell_2$ vector-norm.

\subsection{Definition}
\label{subsec:TFdef}

Trend filtering can be viewed as a blending of the strengths of variable-knot regression splines (local adaptivity and interpretability) and the strengths of smoothing splines (simplicity and speed). Mathematically, this is achieved by choosing an appropriate set of basis functions and penalizing the least-squares problem with an $\ell_1$ norm on the basis coefficients (sum of absolute values), instead of the $\ell_0$ norm of variable-knot regression splines (sum of indicator functions) or the $\ell_2$ norm of smoothing splines (sum of squares). 

This section is primarily summarized from \cite{Tibs} and \cite{pmlr-v32-wange14}. Let the inputs be ordered with respect to the index, i.e. $t_1<\cdots <t_n$. For the sake of simplicity, we consider the case when the inputs $t_1,\dots,t_n \in (a,b)$ are equally spaced with $\Delta t = t_{i+1}-t_i$. See the aforementioned papers for the generalized definition of trend filtering to unequally spaced inputs.

\begin{figure*}
\centering
\includegraphics[width = \textwidth]{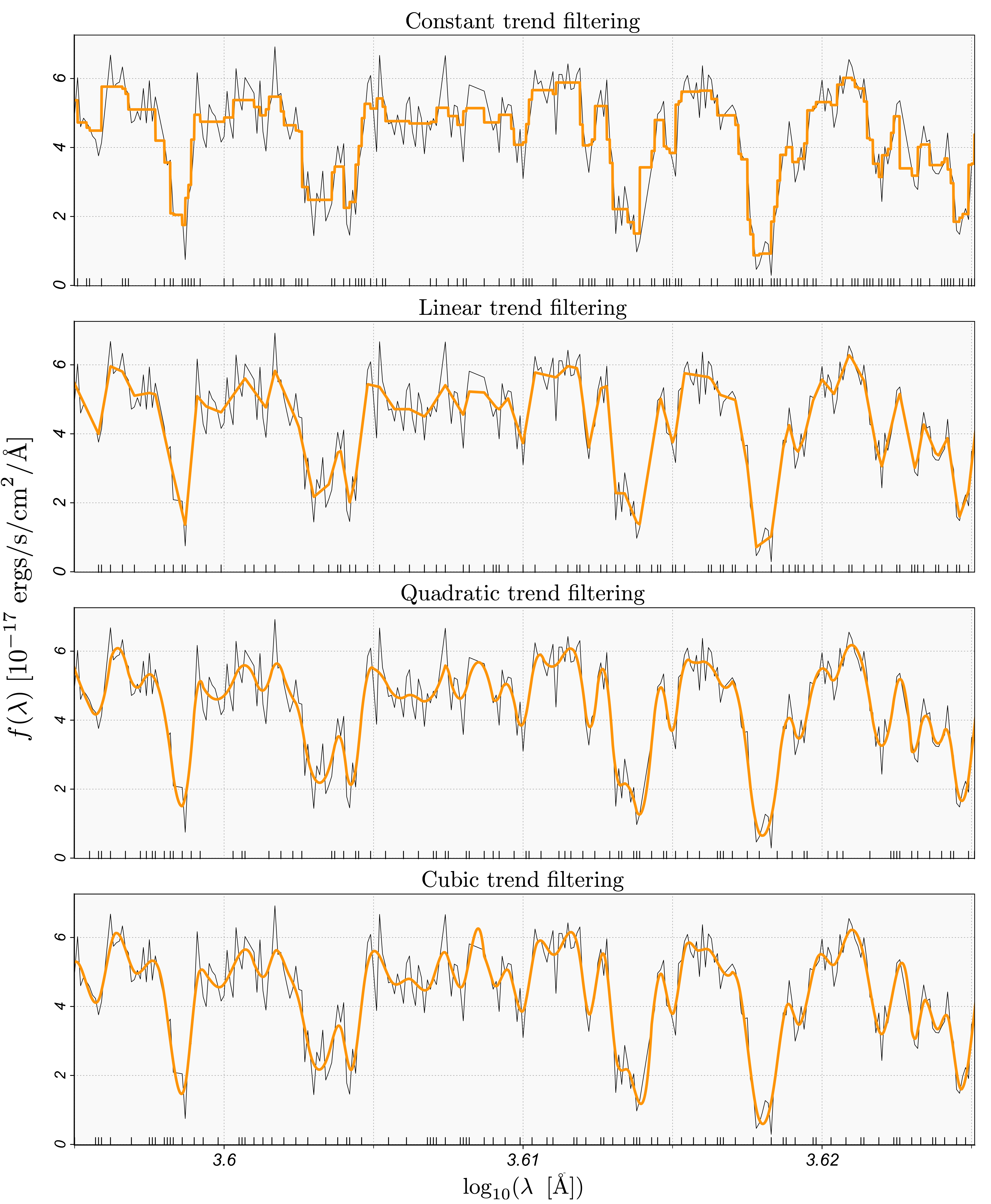}
\caption{Piecewise polynomials with adaptively-chosen knots produced by trend filtering. From top to bottom, we show trend filtering estimates of orders $k=0,1,2$ and $3$, which take the form of piecewise constant, piecewise linear, piecewise quadratic, and piecewise cubic polynomials, respectively. The adaptively-chosen knots of each piecewise polynomial are indicated by the tick marks along the horizontal axes. The constant trend filtering estimate is discontinuous at the knots, but we interpolate here for visual purposes. The data set is taken from the Lyman-$\alpha$ forest of a mock quasar spectrum (\citealt{Bautista}), sampled in logarithmic-angstrom space. We study this phenomenon in detail in Paper~II.}
\label{sample_predictions}
\end{figure*}

For any given integer $k\geq0$, the $k$th order trend filtering estimate is a piecewise polynomial of degree $k$ with knots \emph{automatically selected} at a sparse subset of the observed inputs $t_1,\dots,t_n$. In Figure \ref{sample_predictions}, we provide an example of a trend-filtered data set for orders $k=0,1,2,$ and $3$. Specifically, the panels of the figure respectively display piecewise constant, piecewise linear, piecewise quadratic, and piecewise cubic fits to the data with the automatically-selected knots indicated by the tick marks on the horizontal axes. Constant trend filtering is equivalent to total variation denoising (\citealt{Rudin}), as well as special forms of the fused lasso of \cite{RobTibs} and the variable fusion estimator of \cite{variable_fusion}. Linear trend filtering was independently proposed by \cite{steidl} and \cite{trendfilter}. Higher-order polynomial trend filtering ($k\geq2$) was developed by \cite{genlasso} and \cite{Tibs}. In the Figure \ref{sample_predictions} example, the quadratic and cubic trend filtering estimates are nearly visually indistinguishable, and this is true in general. Although, as we see here, trend filtering estimates of different orders typically select different sets of knots.

Like the spline methods discussed in Section~\ref{subsec:related}, for any order $k\geq0$, the trend filtering estimator has a basis representation\begin{equation}
m(t) = \sum_{j=1}^{n} \beta_j h_j(t),
\end{equation}
but, here, the trend filtering basis $\{h_1,\dots,h_n\}$ is the \emph{falling factorial} basis, which is defined as
\begin{equation}
h_j(t) = \begin{cases}
\prod_{i=1}^{j-1} (t-t_{i}) & j\leq k+1, \\
\prod_{i=1}^{j-1} (t-t_{j-k-1+i})\cdot \mathbbm{1}\{t\geq t_{j-1}\} & j\geq k+2.
\end{cases}
\end{equation}
Like the truncated power basis, the first $k+1$ basis functions span the space of global $k$th order polynomials and the rest of the basis adds the piecewise polynomial structure. However, the knot-producing basis functions of the falling factorial basis $h_j$, $j\geq k+2$ have small discontinuities in their $j$th order derivatives at the knots for all $j=1,\dots,k-1$, and therefore for orders $k\geq 2$ the trend filtering estimate is \emph{close to}, but not quite a spline. The discontinuities are small enough, however, that the trend filtering estimate defined through the falling factorial basis representation is visually indistinguishable from the analogous spline produced by the truncated power basis (see \citealt{Tibs} and \citealt{pmlr-v32-wange14}). The advantage of utilizing the falling factorial basis in this context instead of the truncated power basis (or the B-spline basis) comes in the form of significant computational speedups, as we detail below. 

Analogous to the continuous smoothing spline problem~(\ref{smoothspline1}), we let $\mathcal{H}_k$ be the space of all functions spanned by the $k$th order falling factorial basis, and pose the trend filtering problem as a least-squares minimization with a derivative-based penalty on the fitted function. In particular, the $k$th order trend filtering estimator is the solution to the problem\begin{equation}
\min_{m \in \mathcal{H}_k} \quad  \sum_{i=1}^{n}\big(f(t_i)- m(t_i)\big)^2 + \gamma \cdot \text{TV}(m^{(k)}), \label{TF2}
\end{equation}
where $m^{(k)}$ is the $k$th derivative of $m$, $\text{TV}(m^{(k)})$ is the \emph{total variation} of $m^{(k)}$, and $\gamma>0$ is the model hyperparameter that controls the smoothness of the fit. When $m^{(k)}$ is differentiable everywhere in its domain, the penalty term simplifies to
\begin{equation}
\text{TV}(m^{(k)}) = \int_{a}^{b}|m^{(k+1)}(t)|dt. \label{TV}
\end{equation}
Avoiding the technical generalized definition of total variation \cite[see, e.g.,][]{Tibs}, we can simply think of $\text{TV}(\cdot)$ as a generalized $L_1$ norm\footnote{We use the upper-case notation $L_p$, $p=1,2$ for the $p$-norm of a continuous function, and $\ell_p$, $p=0,1,2$ for the $p$-norm of a vector.} for our piecewise polynomials that possess small discontinuities in the derivatives. Again referring back to the smoothing spline problem~(\ref{smoothspline1}), definitions (\ref{TF2}) and (\ref{TV}) reveal that trend filtering can be thought of as an $L_1$ analog of the ($L_2$-penalized) smoothing spline problem. Moreover, note that unlike smoothing splines, trend filtering can produce piecewise polynomials of all orders $k\geq0$.

Replacing $m$ with its basis representation, i.e. $m(t) = \sum_j \beta_jh_j(t)$, yields the equivalent finite-dimensional trend filtering minimization problem\footnote{This may be recognized as a lasso regression (\citealt{lasso}), with the features being the falling factorial basis functions.}:
\begin{equation}
\min_{\{\beta_j\}} \quad \sum_{i=1}^{n}\Bigg(f(t_i)- \sum_{j=1}^n\beta_jh_j(t_i)\Bigg)^2 + \gamma \cdot k!\cdot \Delta t^k \sum_{j=k+2}^{n}|\beta_j|. \label{lasso}
\end{equation}
The terms $k!$ and $\Delta t^k$ are constants and can therefore be ignored by absorbing them into the hyperparameter $\gamma$. Visual inspection of (\ref{lasso}) reveals that trend filtering is also analogous to the variable-knot regression spline problem~(\ref{variableknot})---namely, by replacing the $\ell_0$ norm on the basis coefficients with an $\ell_1$ norm. The advantage here is that the problem is now strictly convex and can be efficiently solved by various convex optimization algorithms. Furthermore, the $\ell_1$ penalty still yields a sparse solution (i.e. many $\beta_j = 0$), which provides the automatic knot-selection property. Letting $\what{\beta}_1,\dots,\what{\beta}_n$ denote the solution to (\ref{lasso}) for a particular choice of $\gamma>0$, the trend filtering estimate is then given by
\begin{equation}
\what{f}_0(t; \gamma) = \sum_{j=1}^{n}\what{\beta}_jh_j(t), \label{basisdecomp}
\end{equation}
with the automatically-selected knots corresponding to the basis functions with $\what{\beta}_j \neq 0$, $j\geq k+1$.

\begin{table*}
\center
\begin{tabular}{l|l|c|c} 
	\midrule\midrule
	& Method & \multicolumn{1}{p{2.75cm}|}{\centering Computational \\ Complexity} & \multicolumn{1}{p{2.1cm}}{\centering Hyperparameters \\ to estimate}  \\
	\midrule\midrule 
	\multirow{3}{*}{Locally-adaptive}& Wavelets & $\mathcal{O}(n)$ & 1\\ 
	& {\bf Trend filtering} & $\boldsymbol{\mathcal{O}(n^{1.5})}$ & {\bf 1} \\
	& Variable-knot regression splines & $\mathcal{O}(n\cdot \binom{n}{p})$ & 1 \\
	\midrule
	\multirow{5}{*}{Non-adaptive} & Uniform-knot regression splines & $\mathcal{O}(n)$ & 1 \\
	& Smoothing splines & $\mathcal{O}(n)$ & 1 \\
	& Kernel smoothers & $\mathcal{O}(n^2)$ & 1 \\ 
	& LOESS & $\mathcal{O}(n^2)$ & 1 \\ 
	& Gaussian process regression & $\mathcal{O}(n^3)$ & 3+ 
	\\ \midrule\midrule
\end{tabular} 
\caption{Comparison of computational costs associated with popular one-dimensional nonparametric regression methods. The computational complexity column states how the number of elementary operations necessary to obtain the fitted values of each estimator (i.e. the estimator evaluated at the observed inputs) scales with the sample size $n$. For trend filtering, the $\mathcal{O}(n^{1.5})$ complexity represents the worst-case complexity of the \protect\cite{Ramdas} convex optimization algorithm. In most practical settings the actual complexity of this algorithm is close to $\mathcal{O}(n)$. Variable-knot regression splines require a (nonconvex)  exhaustive combinatorial search over the set of possible knots and the complexity therefore includes a binomial coefficient term $\binom{n}{p} = n!/(n!(n-p)!)$, where $p$ is the number of knots in the spline. The remaining methods are explicitly solvable and the stated complexity represents the cost of an exact calculation. The $\mathcal{O}(n)$ complexity of wavelets relies on restrictive sampling assumptions (e.g., equally-spaced inputs, sample size equal to a power of two). The stated computational complexity of all methods represents the cost of a single model fit and does not include the cost of hyperparameter tuning. Gaussian process regression suffers from the most additional overhead in this regard because of the (often) large number of hyperparameters used to parametrize the covariance function (e.g., shape, range, marginal variance, noise variance). Each of the non-adaptive methods (linear smoothers) can be made to be locally adaptive (e.g., by locally varying the hyperparameters of the model), but at the expense of greatly increasing the dimensionality of the hyperparameter space to be searched over.} \label{complexity_comparison}
\end{table*}

The advantage of utilizing the falling factorial basis is found by reparametrizing the problem~(\ref{lasso}) into an optimization over the fitted values $m(t_1),\dots,m(t_n)$. The problem then reduces to
\begin{equation}
\min_{\{m(t_i)\}} \quad \sum_{i=1}^{n}\big(f(t_i)-m(t_i)\big)^2 + \gamma \sum_{i=1}^{n-k-1}|\Delta^{(k+1)}m(t_i)|\cdot \Delta t \label{TF}
\end{equation}
where $\Delta^{(k+1)}m(t_i)$ can be viewed as a discrete approximation of the $(k+1)$st derivative of $m$ at $t_i$. For $k=0$ the discrete derivatives are
\begin{equation}
\Delta^{(1)}m(t_i) = \frac{m(t_{i+1}) - m(t_i)}{\Delta t},
\end{equation}
and then can be defined recursively for $k\geq1$:
\begin{equation}
\Delta^{(k+1)}m(t_i) = \frac{\Delta^{(k)}m(t_{i+1}) - \Delta^{(k)}m(t_i)}{\Delta t}.
\end{equation}
The penalty term in (\ref{TF}) can be viewed as a Riemann-like discrete approximation of the integral in (\ref{TV}). Because of the choice of basis, the problem has reduced to a simple generalized lasso problem \cite[][]{genlasso,genlasso2} with an identity predictor matrix and a banded\footnote{A banded matrix only contains nonzero elements in the main diagonal and zero or more diagonals on either side.} penalty matrix. This special structure allows the solution to be computed very efficiently and with a nearly linear time scaling---i.e. $\mathcal{O}(n)$ elementary operations---via the specialized alternating direction method of multipliers (ADMM) algorithm of \cite{Ramdas}. This algorithm has a linear complexity per iteration, so the overall complexity is $\mathcal{O}(nr)$ where $r$ is the number of iterations necessary to converge to the solution. In the worst case scenario $r\sim n^{1/2}$, so the worst-case overall complexity is $\mathcal{O}(n^{1.5})$. In practice, the computations of the specialized trend filtering optimization algorithm are highly efficient and scale to massive data sets, e.g. handling data sets with $n\gtrsim10^7$ within a few minutes. See \cite{Ramdas} for more rigorous timing results. The practical and scalable computational speed further illustrates the value of trend filtering to astronomy, as it is readily compatible with the large-scale analysis of one-dimensional data sets that has become increasingly ubiquitous in large sky surveys. We show a comparison in Table~\ref{complexity_comparison} of the computational costs associated with trend filtering and other popular one-dimensional nonparametric methods.

Given the trend filtering fitted values obtained by the optimization~(\ref{TF}) the full continuous-time representation of the trend filtering estimate follows by inverting the parametrization back to the basis function coefficients and plugging them into the basis representation (\ref{basisdecomp}).

\subsection{Extension to heteroskedastic weighting}
\label{subsec:hetero}

Thus far we have considered the simple case where the observations are treated as equally-weighted in the cost functional (\ref{TF}). Recall from (\ref{observational_model2}) the observational data generating process and define $\sigma_i^2 = \text{Var}(\epsilon_i)$ to be the noise level---the (typically heteroskedastic) uncertainty in the measurements that arises from instrumental errors and removal of systematic effects. When estimates for $\sigma_i^2$, $i=1,\dots,n$ accompany the observations, as they often do, they can be used to weight the observations to yield a more efficient statistical estimator (i.e. smaller mean-squared error). The error-weighted trend filtering estimator is the solution to the following minimization problem:
\begin{equation}
\min_{\{m(t_i)\}} \quad \sum_{i=1}^{n}\big(f(t_i)-m(t_i)\big)^2w_i + \gamma \sum_{i=1}^{n-k-1}|\Delta^{(k+1)}m(t_i)|\cdot \Delta t,\label{TF3}
\end{equation}
where the optimal choice of weights is $w_i = \sigma_i^{-2}$, $i=1,\dots,n$. Much of the publically available software for trend filtering allows for a heteroskedastic weighting scheme (see Section \ref{subsec:software}).

\subsection{Software}
\label{subsec:software}

Trend filtering software is available online across various platforms. For the specialized ADMM algorithm of \cite{Ramdas} that we utilize in this work, implementations are available in \textsf{R} and \textsf{C} \cite[][]{glmgen}, as well as \textsf{Julia} \cite[][]{Kornblith_software}. \textsf{Matlab} and \textsf{Python} implementations are available for the primal-dual interior point method of \cite{trendfilter}, but only for equally-weighted linear trend filtering \cite[][]{Kim_software,cvxpy}. We provide links to our recommended implementations in Table~\ref{software}. Note that in all software implementations the trend filtering hyperparameter is called $\lambda$ instead of $\gamma$, which we use here to avoid ambiguity with the notation for wavelength in our spectroscopic analyses in Paper~II.

\begin{table}
\center
\begin{tabular}{l|l} 
	\midrule\midrule
	Language & Recommended implementation  \\
	\midrule\midrule 
	R & \MYhref[black]{https://github.com/glmgen/glmgen}{\texttt{github.com/glmgen}} \\
	C & \MYhref[black]{https://github.com/glmgen/glmgen}{\texttt{github.com/glmgen}} \\
	Python & \MYhref[black]{https://www.cvxpy.org/examples/applications/l1_trend_filter.html}{\texttt{cvxpy.org}} \\
	Matlab & \url{http://stanford.edu/~boyd/l1_tf} \hspace{2ex} \\
	Julia & \MYhref[black]{http://github.com/JuliaStats/Lasso.jl}{\texttt{github.com/JuliaStats/Lasso.jl}} 
	\\ \midrule\midrule
\end{tabular} 
\caption{Recommended implementations for trend filtering in various programming languages. See Section~\ref{subsec:software} for details. We provide supplementary \textsf{R} code at \MYhref[black]{http://github.com/capolitsch/trendfilteringSupp}{\texttt{github.com/capolitsch/trendfilteringSupp}} for selecting the hyperparameter via minimization of Stein's unbiased risk estimate (see Section~\ref{subsec:complexity}) and various bootstrap methods for uncertainty quantification (see Section~\ref{subsec:uc}). Our implementations are built on top of the \texttt{glmgen} \textsf{R} package of \protect\cite{glmgen}.} \label{software}
\end{table}

\subsection{Choosing the hyperparameter}
\label{subsec:complexity}
The choice of the piecewise polynomial order $k$ generally has minimal effect on the performance of the trend filtering estimator in terms of mean-squared error and therefore can be treated as an \emph{a priori} aesthetic choice based on how much smoothness is desired or believed to be present. For example, we use $k=2$ (quadratic trend filtering) throughout our analyses in Paper~II so that the fitted curves are smooth, i.e. differentiable everywhere. 

Given the choice of $k$, the hyperparameter $\gamma>0$ is used to tune the complexity (i.e. the wiggliness) of the trend filtering estimate by weighting the tradeoff between the complexity of the estimate and the size of the squared residuals. Obtaining an accurate estimate is therefore intrinsically tied to finding an optimal choice of $\gamma$. The selection of $\gamma$ is typically done by minimizing an estimate of the mean-squared prediction error (MSPE) of the trend filtering estimator. Here, there are two different notions of error to consider, namely, \emph{fixed-input} error and \emph{random-input} error. As the names suggest, the distinction between which type of error to consider is made based on how the inputs are sampled. As a general rule-of-thumb, we recommend optimizing with respect to fixed-input error when the inputs are regularly-sampled and optimizing with respect to random-input error on irregularly-sampled data.

Recall the DGP stated in (\ref{observational_model2}) and let it be denoted by $Q$ so that $\mathbb{E}_Q[\cdot]$ is the mathematical expectation with respect to the randomness of the DGP. Further, let $\sigma_i^2 = \text{Var}(\epsilon_i)$. The fixed-input MSPE is given by
\begin{align}
R(\gamma) &= \frac{1}{n}\sum_{i=1}^{n}\mathbb{E}_{Q}\Big[\big(f(t_i) - \widehat{f}_0(t_i;\gamma)\big)^2\;\Big|\;t_1,\dots,t_n\Big] \label{fixed_error} \\
&= \frac{1}{n}\sum_{i=1}^{n}\Big(\mathbb{E}_{Q}\Big[\big(f_0(t_i) - \widehat{f}_0(t_i;\gamma)\big)^2\;\Big|\;t_1,\dots,t_n\Big] + \sigma_i^2\Big)
\end{align}
and the random-input MSPE is given by
\begin{equation}
\widetilde{R}(\gamma) = \mathbb{E}_{Q}\Big[\big(f(t) - \widehat{f}_0(t;\gamma)\big)^2\Big], \label{random_error}
\end{equation}
where, in the latter, $t$ is considered to be a random component of the DGP with a marginal probability density $p_t(t)$ supported on the observed input interval. In each case, the theoretically optimal choice of $\gamma$ is defined as the minimizer of the respective choice of error. Empirically, we estimate the theoretically optimal choice of $\gamma$ by minimizing an estimate of (\ref{fixed_error}) or (\ref{random_error}). For fixed-input error we recommend Stein's unbiased risk estimate (SURE; \citealt{stein1981,edf}) and for random-input error we recommend $K$-fold cross validation with $K=10$. We elaborate on SURE here and refer the reader to \cite{Wasserman_2003} for $K$-fold cross validation. 

The SURE formula provides an unbiased estimate of the fixed-input MSPE of a statistical estimator:
\begin{align}
\what{R}_0(\gamma) &= \frac{1}{n}\sum_{i=1}^{n}\big(f(t_i) - \what{f}_0(t_i; \gamma)\big)^2 + \frac{2\overline{\sigma}^{2}\text{df}(\what{f}_0)}{n}, \label{Cp}
\end{align}
where $\overline{\sigma}^{2} = n^{-1}\sumin \sigma_i^2$ and $\text{df}(\what{f}_0)$ is defined in (\ref{edf}). A formula for the effective degrees of freedom of the trend filtering estimator is available via the generalized lasso results of \cite{dflasso}; namely,
\begin{align}
\text{df}(\what{f}_0) &= \mathbb{E}[\text{number of knots in $\what{f}_0$}] + k + 1.  \label{edf2}
\end{align}
We then obtain our hyperparameter estimate $\what{\gamma}$ by minimizing the following plug-in estimate for (\ref{Cp}):
\begin{equation}
\what{R}(\gamma) = \frac{1}{n}\sum_{i=1}^{n}\big(f(t_i) - \what{f}_0(t_i; \gamma)\big)^2 + \frac{2\what{\overline{\sigma}}^{2}\what{\text{df}}(\what{f}_0)}{n}, \label{SURE}
\end{equation}
where $\what{\text{df}}$ is the estimate for the effective degrees of freedom that is obtained by replacing the expectation in (\ref{edf2}) with the observed number of knots, and $\what{\overline{\sigma}}^2$ is an estimate of $\overline{\sigma}^2$. If a reliable estimate of $\overline{\sigma}^2$ is not available \emph{a priori}, a data-driven estimate can be constructed (see, e.g., \citealt{Wasserman}). We provide a supplementary \textsf{R} package on the corresponding author's GitHub page\footnote{\url{https://github.com/capolitsch/trendfilteringSupp}} for implementing SURE with trend filtering. The package is built on top of the\hspace{1ex}\texttt{glmgen}\hspace{1ex}\textsf{R} package of \cite{glmgen}, which already includes an implementation of $K$-fold cross validation.

Because of the existence of the degrees of freedom expression (\ref{edf2}), trend filtering is also compatible with reduced chi-squared model assessment and comparison procedures under a Gaussian noise assumption \cite[][]{Pearson,cochran1952}.


\subsection{Uncertainty quantification}
\label{subsec:uc}

\subsubsection{Frequentist}

Frequentist uncertainty quantification for trend filtering follows by studying the sampling distribution of the estimator that arises from the randomness of the observational data generating process (DGP). In particular, most of the uncertainty in the estimates is captured by studying the variability of the estimator with respect to the DGP. We advise three different bootstrap methods \cite[][]{efron1979} for estimating the variability of the trend filtering estimator, with each method corresponding to a distinct analysis setting. Here, we emphasize the terminology \emph{variability}---as opposed to the variance of the trend filtering estimator---since, by construction, as a nonlinear function of the observed data, the trend filtering estimator has a non-Gaussian sampling distribution even when the observational noise is Gaussian. For that reason, each of our recommended bootstrap approaches is based on computing sample quantiles (instead of pairing standard errors with Gaussian quantiles).

We restate the assumed DGP here for clarity:
\begin{equation}
f(t_i) = f_0(t_i) + \epsilon_i,  \hfill t_1,\dots,t_n\in(a,b)
\end{equation} 
where $\mathbb{E}[\epsilon_i] = 0$. We make the further assumption that the errors $\epsilon_1,\dots,\epsilon_n$ are independent\footnote{If nontrivial autocorrelation exists in the noise then a block bootstrap \cite[][]{kunsch1989} will yield a better approximation of the trend filtering variability than the bootstrap implementations we discuss.}. The three distinct settings we consider are:
\begin{enumerate}[leftmargin=*,labelindent=10pt]
\item[\bf S1.] The inputs are irregularly sampled
\item[\bf S2.] The inputs are regularly sampled and the noise distribution is known
\item[\bf S3.] The inputs are regularly sampled and the noise distribution is unknown
\end{enumerate}
The corresponding bootstrap methods are detailed in Algorithm~\ref{Alg1} \cite[nonparametric bootstrap;][]{efron1979}, Algorithm~\ref{Alg2} \cite[parametric bootstrap;][]{bootstrap}, and Algorithm~\ref{Alg3} \cite[wild bootstrap;][]{Wu,Liu,mammen1993}, respectively. We include implementations of each of these algorithms in the \textsf{R} package on our GitHub page.

\begin{algorithm}
\caption{\small Nonparametric bootstrap for random-input uncertainty quantification}
\begin{algorithmic}[1]
\REQUIRE {Training Data $(t_1,f(t_1)),\dots,(t_n,f(t_n))$, hyperparameters $\gamma$ and $k$, prediction input grid $t_1',\dots,t_m'$} 
\FORALL{$b$ in $1:B$}
	\STATE Define a bootstrap sample of size $n$ by resampling the observed pairs with replacement:
	\begin{equation*}
	(t_1^{*}, f_b^{*}(t_1^{*})),\dots,(t_n^{*}, f_b^{*}(t_n^{*}))
	\end{equation*}
	\STATE Let $\what{f}_b^{*}(t_1'), \dots,\what{f}_b^{*}(t_m')$ denote the trend filtering estimate fit on the bootstrap sample and evaluated on the prediction grid $t_1',\dots,t_m'$
\ENDFOR \\
\hspace{-3.75ex}{\bf Output:} The full trend filtering bootstrap ensemble \\ \hspace{6ex} $\{\what{f}_b^{*}(t_i')\}_{\substack{i=1,\dots,m \\ b=1,\dots,B}}$
\end{algorithmic} \label{Alg1}
\end{algorithm}

\begin{algorithm}
\caption{\small Parametric bootstrap for fixed-input uncertainty quantification (when noise distribution $\epsilon_i\sim Q_i$ is known \emph{a priori})}
\begin{algorithmic}[1]
\REQUIRE {Training Data $(t_1,f(t_1)),\dots,(t_n,f(t_n))$, hyperparameters $\gamma$ and $k$, assumed noise distribution $\epsilon_i \sim Q_i$, prediction input grid $t_1',\dots,t_m'$}
\STATE Compute the trend filtering point estimate at the observed inputs:
\begin{equation*}
(t_1,\what{f}_0(t_1)),\dots, (t_n,\what{f}_0(t_n))
\end{equation*}
\FORALL{$b$ in $1:B$}
	\STATE Define a bootstrap sample by sampling from the assumed noise distribution:
\begin{equation*}
f_b^{*}(t_i) = \what{f}_0(t_i) + \epsilon_i^{*} \quad \quad \text{where }\epsilon_i^{*} \sim Q_i, \quad i=1,\dots,n
\end{equation*}
\STATE Let $f_b^{*}(t_1'), \dots,f_b^{*}(t_m')$ denote the trend filtering estimate fit on the bootstrap sample and evaluated on the prediction grid $t_1',\dots,t_m'$
\ENDFOR \\
\hspace{-3.75ex}{\bf Output:} The full trend filtering bootstrap ensemble \\ \hspace{6ex} $\{\what{f}_b^{*}(t_i')\}_{\substack{i=1,\dots,m \\ b=1,\dots,B}}$
\end{algorithmic} \label{Alg2}
\end{algorithm}

\begin{algorithm}
\caption{\small Wild bootstrap for fixed-input uncertainty quantification (when noise distribution is not known \emph{a priori})}
\begin{algorithmic}[1]
\REQUIRE {Training Data $(t_1,f(t_1)),\dots,(t_n,f(t_n))$, hyperparameters $\gamma$ and $k$, prediction input grid $t_1',\dots,t_m'$} 
\STATE Compute the trend filtering point estimate at the observed inputs:
\begin{equation*}
(t_1,\what{f}_0(t_1)),\dots, (t_n,\what{f}_0(t_n))
\end{equation*}
\STATE Let $\what{\epsilon}_i = f(t_i) - \what{f}_0(t_i)$, $i=1,\dots,n$ denote the residuals
\FORALL{$i$}
	\STATE Define a bootstrap sample by sampling from the following distribution:
\begin{equation*}
f_b^{*}(t_i) = \what{f}_0(t_i) + u_i^{*} \quad \quad i=1,\dots,n
\end{equation*}
where
\begin{equation*}
u_i^{*} = \begin{cases}
\what{\epsilon}_i(1 + \sqrt{5})/2 & \text{with probability }(1+\sqrt{5})/(2\sqrt{5}) \\
\what{\epsilon}_i(1 - \sqrt{5})/2 & \text{with probability }(\sqrt{5}-1)/(2\sqrt{5})
\end{cases}
\end{equation*}
\STATE Let $f_b^{*}(t_1'), \dots,f_b^{*}(t_m')$ denote the trend filtering estimate fit on the bootstrap sample and evaluated on the prediction grid $t_1',\dots,t_m'$
\ENDFOR \\
\hspace{-3.75ex}{\bf Output:} The full trend filtering bootstrap ensemble \\ \hspace{6ex} $\{\what{f}_b^{*}(t_i')\}_{\substack{i=1,\dots,m \\ b=1,\dots,B}}$
\end{algorithmic} \label{Alg3}
\end{algorithm}

Given the full trend filtering bootstrap ensemble provided by the relevant bootstrap algorithm, for any $\alpha\in(0,1)$, a $(1-\alpha)\cdot100\%$ quantile-based pointwise variability band is given by
\begin{equation}
V_{1-\alpha}(t_i') = \Big(\what{f}_{\alpha/2}^{*}(t_i'),\;\what{f}_{1-\alpha/2}^{*}(t_i')\Big), \hfill i = 1,\dots,m \label{variability_band}
\end{equation}
where
\begin{equation}
\what{f}_{\beta}^{*}(t_i') = \inf_{g}\Bigg\{g : \frac{1}{B} \sum_{b=1}^{B} \mathbbm{1}\big\{\what{f}_b^{*}(t_i') \leq g\big\} \geq \beta \Bigg\}, \hfill \beta\in(0,1).
\end{equation}
Analogously, bootstrap sampling distributions and variability intervals for observable parameters of the signal may be studied by deriving a bootstrap parameter estimate from each trend filtering estimate within the bootstrap ensemble. For example, in Paper~II we examine the bootstrap sampling distributions of several observable light-curve parameters of exoplanet transits and supernovae.

\subsubsection{Bayesian}

There is a well-studied connection between $\ell_1$-penalized least-squares regression and a Bayesian framework \cite[see, e.g.,][]{lasso,bayesian_lasso2,bayesian_lasso}. A discussion specific to trend filtering can be found in \cite{faulkner2018}.


\subsection{Relaxed trend filtering}
\label{subsec:relaxed}

We are indebted to Ryan Tibshirani for a private conversation that motivated the discussion in this section. Trend filtering can be generalized to allow for greater flexibility through a technique that we call \emph{relaxed trend filtering}\footnote{We choose this term because the generalization of trend filtering to relaxed trend filtering is analogous to the generalization of the lasso \cite[][]{lasso} to the relaxed lasso \cite[][]{relaxo}.}. Although the traditional trend filtering estimator is already highly flexible, there are certain settings in which the relaxed trend filtering estimator provides nontrivial improvements. In our experience, these typically correspond to settings where the optimally-tuned trend filtering estimator selects very few knots. For example, we use relaxed trend filtering in Paper~II to model the detrended, phase-folded light curve of a Kepler star with a planetary transit event.

The relaxed trend filtering estimate is defined through a two-stage sequential procedure in which the first stage amounts to computing the traditional trend filtering estimate discussed in Section \ref{subsec:TFdef}. Recall the trend filtering minimization problem (\ref{lasso}). For any given order $k\in\{0,1,2,\dots\}$ and hyperparameter $\gamma>0$, let us amend our notation so that 
\begin{equation}
\what{f}_0^{\;TF}(t) = \sum_{j=1}^{n} \what{\beta}_j^{\;TF}h_j(t)
\end{equation}
denotes the basis representation of the traditional trend filtering estimate. Further, define the index set
\begin{equation}
\mathcal{K}_{\gamma} = \Big\{1\leq j\leq n\; | \; \what{\beta}_j^{\;TF} \neq 0\Big\}
\end{equation}
that includes the indices of the non-zero falling factorial basis coefficients for the given choice of $\gamma$. Now let $\what{\beta}_j^{\;OLS}$, $j\in \mathcal{K}_{\gamma}$, denote the solution to the ordinary least-squares (OLS) minimization problem\begin{equation}
\min_{\{\beta_j\}} \quad  \sum_{i=1}^{n}\Bigg(f(t_i)- \sum_{j\in \mathcal{K}_{\gamma}}\beta_jh_j(t_i)\Bigg)^2,
\end{equation}
and define the corresponding OLS estimate as
\begin{equation}
\what{f}_{0}^{\;OLS}(t) = \sum_{j\in \mathcal{K}_{\gamma}}\what{\beta}_j^{\;OLS}h_j(t). \label{OLS}
\end{equation}
That is, the OLS estimate (\ref{OLS}) uses trend filtering to find the knots in the piecewise polynomial, but then uses ordinary least-squares to estimate the reduced set of basis coefficients. The relaxed trend filtering estimate is then defined as a weighted average of the traditional trend filtering estimate and the corresponding OLS estimate:\begin{equation}
\what{f}_0^{\;RTF}(t) = \phi\what{f}_0^{\;TF}(t) + (1-\phi)\what{f}_0^{\;OLS}(t),
\end{equation}
for some choice of relaxation hyperparameter $\phi\in[0,1]$. Relaxed trend filtering is therefore a generalization of trend filtering in the sense that the case $\phi=1$ returns the traditional trend filtering estimate.

In principle, it is preferable to jointly optimize the trend filtering hyperparameter $\gamma$ and the relaxation hyperparameter $\phi$, e.g. via cross validation. However, it often suffices to choose $\gamma$ and $\phi$ sequentially, which in turn adds minimal computational cost on top of the traditional trend filtering procedure. Because of the trivial proximity of the falling factorial basis to the truncated power basis (established in \citealt{Tibs} and \citealt{pmlr-v32-wange14}), it is sufficient to let $\what{f}_0^{\;OLS}$ be the $k$th order regression spline with knots at the input locations selected by the trend filtering estimator. In heteroskedastic settings, as discussed in Section~\ref{subsec:hetero}, a piecewise polynomial or regression spline fit by weighted least-squares should be used in place of the OLS estimate~(\ref{OLS}).

\section{Concluding remarks}

The analysis of one-dimensional data arising from signals possessing varying degrees of smoothness is central to a wide variety of problems in time-domain astronomy and astronomical spectroscopy. Trend filtering is a modern statistical tool that provides a unique combination of (1) statistical optimality for estimating signals with varying degrees of smoothness; (2) natural flexibility for handling practical analysis settings (general sampling designs, heteroskedastic noise distributions, etc.); (3) practical computational speed that scales to massive data sets; and (4) a single model hyperparameter that can be chosen via automatic data-driven methods. Software for trend filtering is freely available online across various platforms and we provide links to our recommendations in Table~\ref{subsec:software}. Additionally, we make supplementary \textsf{R} code available on the corresponding author's GitHub page\footnote{\url{https://github.com/capolitsch/trendfilteringSupp}} for: (1) selecting the trend filtering hyperparameter by minimizing Stein's unbiased risk estimate (see Section \ref{subsec:complexity}); and (2) various bootstrap methods for trend filtering uncertainty quantification (see Section \ref{subsec:uc}).


\section*{ACKNOWLEDGEMENTS}
We gratefully thank Ryan Tibshirani for his inspiration and generous feedback on this topic. This work was partially supported by NASA ATP grant NNX17AK56G, NASA ATP grant 80NSSC18K1015, and NSF grant AST1615940.




\bibliographystyle{mnras}
\addcontentsline{toc}{section}{Bibliography}
\bibliography{mybib}{}
\bibliographystyle{mnras}
\setcitestyle{authoryear,open={[},close={]}}

\bsp	
\label{lastpage}
\end{document}